\documentclass[11pt, onecolumn, superscriptaddress,notitlepage]{revtex4-1}
\usepackage{color}
\usepackage{amsmath,amssymb,graphicx,bm,float}

\newcommand{\ket}[1]{|{#1}\rangle}

\begin{document}
\title{Generation of spectrally factorable photon pairs via multi-order quasi-phase-matched spontaneous parametric downconversion}
\author{Fumihiro Kaneda}
\affiliation{Frontier Research Institute for Interdiciplinary Sciences, Tohoku University, 6-3 Aramaki aza Aoba, Aoba-ku, Sendai 980-8578, Japan}
\affiliation{Research Institute of Electrical Communication, Tohoku University, 2-1-1 Katahira, Sendai, 980-8577, Japan}
\author{Jo Oikawa}
\affiliation{Research Institute of Electrical Communication, Tohoku University, 2-1-1 Katahira, Sendai, 980-8577, Japan}
\author{Masahiro Yabuno}
\affiliation{Advanced ICT Research Institute, National Institute of Information and Communications Technology, 588-2 Iwaoka, Nishi-ku, Kobe, Hyogo 651-2492, Japan}
\author{Fumihiro China}
\affiliation{Advanced ICT Research Institute, National Institute of Information and Communications Technology, 588-2 Iwaoka, Nishi-ku, Kobe, Hyogo 651-2492, Japan}
\author{Shigehito Miki}
\affiliation{Advanced ICT Research Institute, National Institute of Information and Communications Technology, 588-2 Iwaoka, Nishi-ku, Kobe, Hyogo 651-2492, Japan}
\affiliation{Graduate School of Engineering, Kobe University, 1-1 Rokko-dai cho, Nada-ku, Kobe 657-0013, Japan }
\author{Hirotaka Terai}
\affiliation{Advanced ICT Research Institute, National Institute of Information and Communications Technology, 588-2 Iwaoka, Nishi-ku, Kobe, Hyogo 651-2492, Japan}
\author{Yasuyoshi Mitsumori}
\affiliation{Department of Physics, School of Science, Kitasato University, 1-15-1 Kitazato, Minami-ku, Sagamihara, 252-0373, Japan}
\author{Keiichi Edamatsu}
\affiliation{Research Institute of Electrical Communication, Tohoku University, 2-1-1 Katahira, Sendai, 980-8577, Japan}

\begin{abstract}
For advanced quantum information technology, sources of photon pairs in quantum mechanically factorable states are of great importance for realizing high-fidelity photon-photon quantum gate operations. 
Here we experimentally demonstrate a technique to produce spectrally factorable photon pairs utilizing multi-order quasi-phase-matching (QPM) conditions in spontaneous parametric downconversion (SPDC). 
In our scheme, a spatial nonlinearity profile of a nonlinear optical crystal is shaped with current standard poling techniques, and the associated phase-matching function can be approximated to a Gaussian form.  
By the measurement of a phase-matching function and the second-order autocorrelation function, we demonstrate that telecom-band photon pairs produced by our custom-poled crystal are highly factorable with $> 95$\% single-photon purity. 
\end{abstract}

%\date{\today}
\maketitle

\section{Introduction}
Quantum photonics technology plays an important role in various quantum applications such as quantum computing \cite{O'Brien.2009,Zhong.2020}, quantum communication \cite{Gisin.2002, Liao.2017}, quantum metrology \cite{Giovannetti.2011ohj}, and bridging solid and atomic quantum systems at a distance \cite{Awschalom.2021}. 
One of the central requirements in the quantum photonics technology is to produce pure single photons. 
Most multi-photon applications rely on interference of independent single photons \cite{Hong.1987}, and only pure and indistinguishable photons can exhibit perfect interference. 
Photon-pair generation via spontaneous parametric downconversion (SPDC) has been widely accepted in quantum optics and photonic quantum information experiments \cite{Pan.2012}. 
Although a single SPDC source produces a photon pair only probabilistically, it is much easier to implement than single-emitter sources based on semiconductor \cite{Senellart.2017} and atomic \cite{McKeever.2004pnl} systems that require cryogenic and/or ultra-high vacuum systems. 
Moreover, recently demonstrated multiplexing techniques  \cite{Ma.2011tbo,Collins.20133n,Kaneda.2015,Joshi.2018,Kaneda.2019} have overcome the probabilistic nature of SPDC, illustrating the possibility of the pseudo-on-demand generation of single photons.

For multi-photon quantum information applications, a joint state of an SPDC photon pair needs to be a spectrally factorable state, i.e., a product state of individual pure single photons. 
Such high-factorability sources can also be used as single-pass two-mode squeezed light sources, basic resources of a large-scale Gaussian boson sampler \cite{Zhong.2020}. 
A typical SPDC source has a certain amount of spectral correlation between a pair of photons, and thus does not generate spectrally factorable photon pairs. 
For spectral factorability of photon pairs, this type of source requires narrowband spectral filtering that is accompanied with the significant reduction of the source brightness and the probability of the coincident photon-pair collection and detection, i.e., the heralding efficiency. 

In order to mitigate the spectral correlation of SPDC sources without spectral filtering, group-velocity-matching (GVM) condition \cite{Grice.2001,Giovannetti.2002wu,Konig.2004,Shimizu.2009} of SPDC sources with appropriate broadband pump pulses have been demonstrated \cite{Mosley.2008,Evans.2010, Yabuno.2012, Kaneda.2016,Weston.2016,Greganti.2018}. 
However, there remains residual spectral correlation in such a GVM-SPDC source due to the peripheral lobes of the sinc-shaped phase-matching function that is associated with a uniform longitudinal nonlinearity in a nonlinear optical crystal.

For tailoring a spectral shape of SPDC photons, quasi-phase matching (QPM) techniques via modulated periodic inversion of a nonlinear crystal have been demonstrated. 
While a standard QPM crystal with a single poling period is designed to satisfy one target QPM condition, such a modulated-poling crystal can achieve multiple QPMs in a single crystal, enabling to produce a variety of spectral shapes of photon pairs \cite{Nasr.2008, Tanaka.2012, Kaneda.2019rqn}. 
For factorable photon-pair generation, several theoretical \cite{Branczyk.2010, Dixon.2013, Dosseva.2016} and experimental \cite{Chen.20176tc, Graffitti.2018, Pickston.2021} works have demonstrated the usefulness of modulated periodic poling techniques to shape a phase-matching spectrum close to a Gaussian form, ideal for spectral factorability. 
However, most of the demonstrated schemes require high precision poling process and/or complicated numerical design. 
Since multi-photon quantum information applications require many indistinguishable single photons, an efficient and reproducible scheme with standard poling technology is desirable. 

In this paper, we demonstrate the generation of spectrally factorable photon pairs with multi-order QPM conditions.
A higher-order QPM condition is satisfied by a larger poling period of a nonlinear crystal, having a less effective nonlinearity. 
Therefore, a nonlinearity profile of a SPDC crystal can be engineered by utilizing large poling periods dependent on the position of the crystal without high-precision poling techniques. 
Multi-order QPM techniques for tailoring various forms of phase-matching functions has been proposed in \cite{Branczyk.2010}. 
We show that our custom-poled crystal satisfying multi-order QPM and GVM conditions has an approximate Gaussian phase-matching function at the telecom C-band. 
This enables to generate high-factorability photon pairs incorporated with an optimal pump bandwidth. 
We characterize our CPKTP crystal using two different methods, i.e., the measurement of a phase-matching function and the second-order autocorrelation function, which consistently reveal high purity of produced single photons. 
We also discuss the possibility of further improvement to our SPDC source.
%

%\begin{figure}[h!]
%\centering\includegraphics[width=7cm]{osafig1}
%\caption{Sample caption (Fig. 2, \cite{Yelin:03}).}
%\end{figure}

\section{Engineered phase-matching spectrum with multi-order QPM conditions}
\begin{figure}[t!]
\centering\includegraphics[width=1\columnwidth , clip]{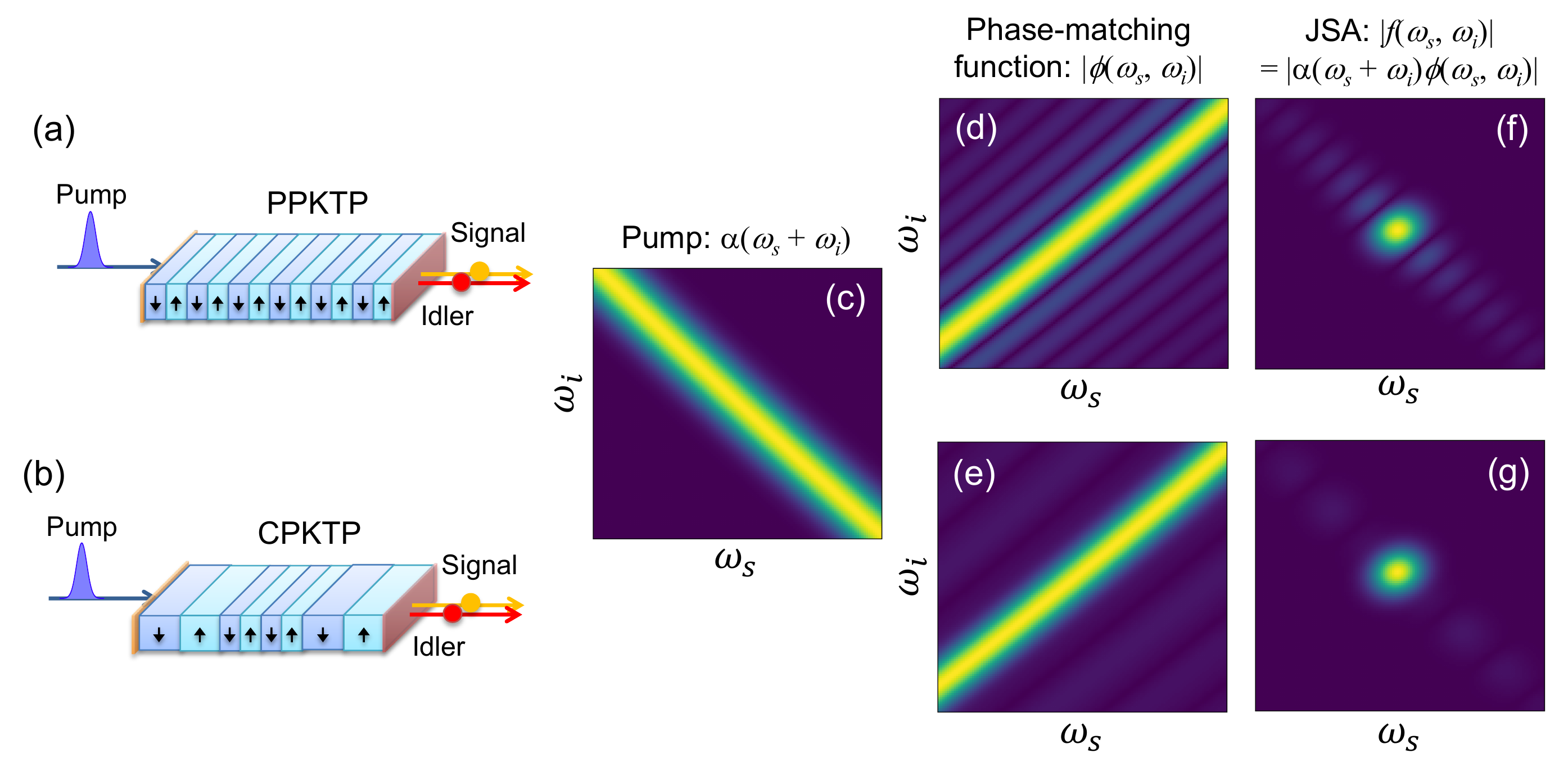}
\caption{Conceptual diagram of telecom-band SPDC sources with (a) periodically-poled potathium titanyl phosphate (PPKTP) crystal and (b) our custom-poled KTP (CPKTP) crystal. (c) Pump envelope function $\alpha (\omega_s +  \omega_i)$. (d-g) Phase-matching function $\phi (\Delta k (\omega_s, \omega_i))$ and joint spectral amplitude (JSA) $f (\omega_s, \omega_i)$ of PPKTP and CPKTP crystals. By utilizing multi-order quasi-phase-matching conditions (QPMs) for different regions of a crystal, $\phi (\Delta k (\omega_s, \omega_i))$ of our CPKTP crystal has an approximated Gaussian form enabling to produce a highly factorable JSA. }
\label{concept}
\end{figure}
Conceptual diagrams of our scheme are depicted in Fig. \ref{concept}. 
SPDC is a nonlinear optical process, where one high-frequency (pump) photon is split into two low-frequency (signal and idler) photons. 
With the plane-wave approximation, a SPDC two-photon joint spectral state is given by

\begin{align}
\label{statevector}
 \ket{\psi_{si}} =  \iint d\omega_s d\omega_i  f(\omega_s, \omega_i ) \ket{\omega_s, \omega_i}, 
%  \ket{\psi_{si}} =  \int d\omega_s d\omega_i  \alpha (\omega_s, \omega_i ) \phi(\Delta k L) \ket{\omega_s, \omega_i}. 
\end{align}
Here, $f (\omega_s, \omega_i )$ is the joint spectral amplitude (JSA), and $\ket{\omega_s, \omega_i}$ denotes a photon-pair state with signal and idler frequencies $\omega_s$ and $\omega_i$, respectively. 
For a collinear SPDC, the JSA is well approximated by 

\begin{align}
\label{JSA}
f(\omega_s, \omega_i ) \simeq  \alpha (\omega_s + \omega_i ) \phi(\Delta k(\omega_s, \omega_i)),
\end{align}
Here, $\alpha(\omega_s + \omega_i )$ is the pump envelope function. 
For a pump pulse with a Gaussian spectral distribution (that is a good approximation of a mode-locked laser pulse) can be described as 

\begin{align}
\label{pump}
\alpha (\omega_s + \omega_i ) &= \exp \left[ \frac{(\omega_s + \omega_i -\omega_{p0} )^2}{\sigma_p^2} \right]  \\
                                                  &= \exp \left[ \frac{(\Omega_s + \Omega_i )^2}{\sigma_p^2} \right],
%  \ket{\psi_{si}} =  \int d\omega_s d\omega_i  \alpha (\omega_s, \omega_i ) \phi(\Delta k L) \ket{\omega_s, \omega_i}. 
\end{align}
where $\Omega_{s(i)} = \omega_{s(i)} - \omega_{s0(i0)}$ is the frequency detuning from the signal (idler) central frequency $\omega_{s0(i0)}$. 
The pump central frequency and bandwidth are respectively denoted by $\omega_{p0} = \omega_{s0} + \omega_{i0}$ and $\sigma_p$. 
As shown in Eq. \ref{pump} and Fig. \ref{concept} (b,e), $\alpha (\omega_s + \omega_i)$ has a maximum value when energy-conservation condition is satisfied: $\Omega_{s} + \Omega_i  = 0$.
The phase-matching function $\phi(\Delta k (\omega_s, \omega_i))$ is given by

\begin{align}
\label{PMF}
\phi(\Delta k) =  \int_{-\infty}^{\infty}  dz \chi^{(2)}(z) e^{i\Delta k z}, 
\end{align}
which is Fourier transform of a longitudinal nonlinearity profile $\chi^{(2)}(z)$. 
The phase mismatch $\Delta k = k_p (\omega_p)- k_s(\omega_s) -k_i(\omega_i)$ represents the difference of the wavenumber $k_x$ of the pump ($x =p$), signal ($x =s$), and ilder ($x =i$) modes. 
Thus, the phase-matching function is essentially determined by the nonlinearity profile and the phase mismatch.

In order to generate a highly factorable JSA, the phase-matching function required to be a positively correlated ($\Omega_i/\Omega_s > 0$) Gaussian function, since the pump envelope function has a negatively correlated ($\Omega_i/\Omega_s = -1$) Gaussian form.  
A widely accepted method for producing a positively correlated phase-matching function is to utilize a Type-II QPM condition in a periodically-poled potassium titanyl phosphate (PPKTP) crystal \cite{Evans.2010, Yabuno.2012,Weston.2016,Greganti.2018}, as shown in Fig. \ref{concept} (a). 
In this configuration, $\Delta k $ is well approximated by the first-order dispersion: $\Delta k(\Omega_s, \Omega_i ) \simeq k'_p (\Omega_s + \Omega_i )  - k'_s\Omega_s  - k'_i \Omega_i$, where $k'_x = \frac{dk_x}{d\omega_x}|_{\omega_{x0}}$ is the inverse group velocity at the photon's central frequency $\omega_{x0}$. 
In the case that the signal and idler photons are produced at a telecom band ($\sim$ 1500-1600 nm), this SPDC process satisfies a GVM condition ($k'_p - k'_s = - k'_p + k'_i$), which result in $\phi(\Delta k (\omega_s, \omega_i))$ with the positive spectral correlation $\Omega_i/\Omega_s = - (k'_p-k'_s)/(k'_p-k'_i) = 1$, as shown in Fig. \ref{concept} (d). 
However, such a PPKTP crystal with a single poling period has a uniform nonlinearity profile, associated with a sinc-shaped phase-matching function.
Therefore, due to peripheral lobes of the sinc distribution, the resulting JSA has a limited spectral factorability and a single-photon purity, as shown in Fig. \ref{concept} (f). 
\begin{figure}[t!]
\centering\includegraphics[width=1\columnwidth , clip]{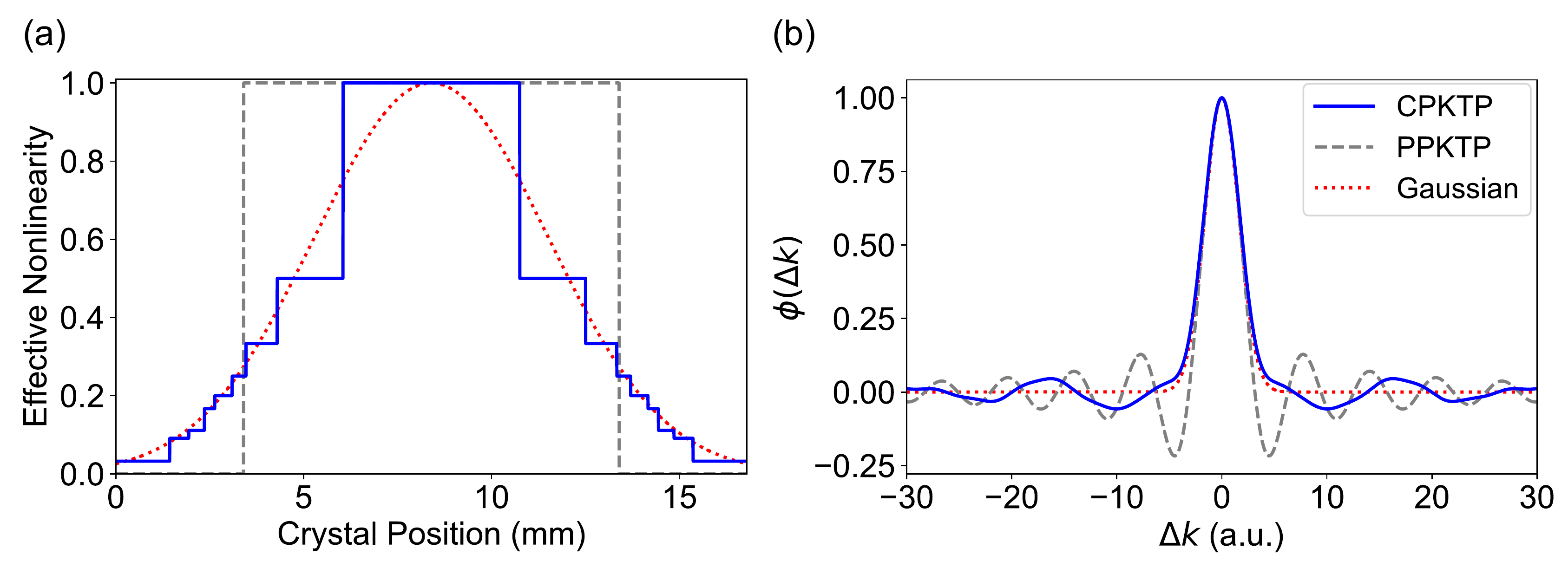}
\caption{(a) Nonlinearity profile and (b) phase-matching function of our designed CPKTP crystal with multi-order QPM conditions (solid line). The nonlinearity profile is approximated to a Gaussian function (dotted line) with 8 mm full width at half maximum (FWHM). The phase-matching function of the designed CPKTP crystal has highly suppressed peripheral lobes compared to that of a PPKTP crystal ($L = 10$ mm), whose phase-matching bandwidth is the same as the target  Gaussian function. }
\label{design}
\end{figure}

%\begin{table}[t]
%\centering
%\caption{ QPM order $m$ and length $l$ (in unit of the 1st-order poling period $\Lambda = 46.1 \mu$m) of each poling region of the designed CPKTP crystal. }
%\begin{tabular}{ l  ccccccccccccccccc }      \hline 
%                 $m$              &31&11&9&6&  5&4&  3&  2&  1  &  2&  3&  4&  5&6&9&11&31 \\  \hline 
%            $l$ ($\Lambda$)&31 &11&9&6&10&8&18&38&102&38&18&  8&10&6&9&11&31 \\   \hline 
%\end{tabular}
%\label{table}
%\end{table}

In our scheme, as shown in Fig. \ref{concept} (b), we use multiple orders of QPM conditions for shaping $\phi(\Delta k)$ close to a Gaussian form. 
In general, a QPM condition is achieved by the periodic poling inversion of a nonlinear crystal that also periodically flips the sign of a nonlinearity. 
For the poling period and duty cycle of $\Lambda$ and $D$, a nonlinearity profile can be written as the following Fourier components:

\begin{align}
\label{poling}
\chi^{(2)}(z) = \sum_{m = -\infty}^{\infty} \frac{2 }{m \pi} \sin^2 (\pi m D )  \sin \frac{2 \pi m }{\Lambda}z.
\end{align}
With this form, we see that a QPM conditions is satisfied for $\Delta k(\Omega_s, \Omega_i ) = 2\pi m /\Lambda$ to produce SPDC photons. 
%また、ある一次の議事位相整合がラムダで成り立つ時、mラムダの周期はm時のQPMを成立させる。
In other words, for a given phase mismatch, if a poling period $\Lambda$ satisfies a 1st-order QPM condition $\Delta k(\Omega_s, \Omega_i ) = 2\pi /\Lambda$, its integer multiple $m \Lambda$ also satisfies a $m$-th order QPM: $\Delta k(\Omega_s, \Omega_i ) = 2\pi m /m\Lambda$. 
Moreover, an amplitude of a Fourier component, i.e., effective nonlinearity, is decreased as absolute number of harmonic order ($2/m\pi$). 
Although inefficient higher-order QPM is not preferable in many classical optics applications, the $m$-dependent nonlinearity can be utilized for shaping a nonlinearity profile. 
% 

%Figure \ref{design} (a) and Table \ref{table} show a poling design of our custom-poled KTP (CPKTP) crystal. 
Figure \ref{design} (a) shows a designed nonlinearity profile of our custom-poled KTP (CPKTP) crystal. 
We used eight different poling periods to achieve 1st- to 31st-order QPM conditions ($\Lambda = 46.1 \mu$m for the 1st-order QPM) for different regions of the crystal to approximate an effective nonlinearity profile to a Gaussian form with the full width at half maximum (FWHM) of 8 mm. 
An effective nonlinearity for each QPM condition is maximized by using the duty cycles $D = 0.5$ and $D = (m-1)/(m+1)$ for odd- and even-order QPM conditions, respectively. 
(Note that one can use modulated poling duty cycles for engineering a nonlinearity profile, as demonstrated in \cite{Dixon.2013, Chen.20176tc}. However, this approach also modulates central frequencies of SPDC photons and needs a high-precision poling technology.)
A predicted phase-matching function of the CPKTP crystal is shown in Fig. \ref{concept} (e) and Fig. \ref{design} (b). 
We see that our designed phase-matching function has smaller peripheral lobes than those of a 10-mm length PPKTP crystal that has the same phase-matching bandwidth at the main peak.  
As shown in Fig. \ref{concept} (g), this enables to generate a highly factorable JSA with a high single-photon purity (97.5\%), comparable to the ones using more complicated, high spatial-resolution poling techniques \cite{Chen.20176tc, Graffitti.2018, Pickston.2021}. 
We also note that large spectral separation of the peripheral lobes and the main peak in our CPKTP crystal allows us to selectively eliminate the lobes by mild spectral filtering, as will be discussed later.

\section{Experiment}
\begin{figure}[t!]
\centering\includegraphics[width=1\columnwidth , clip]{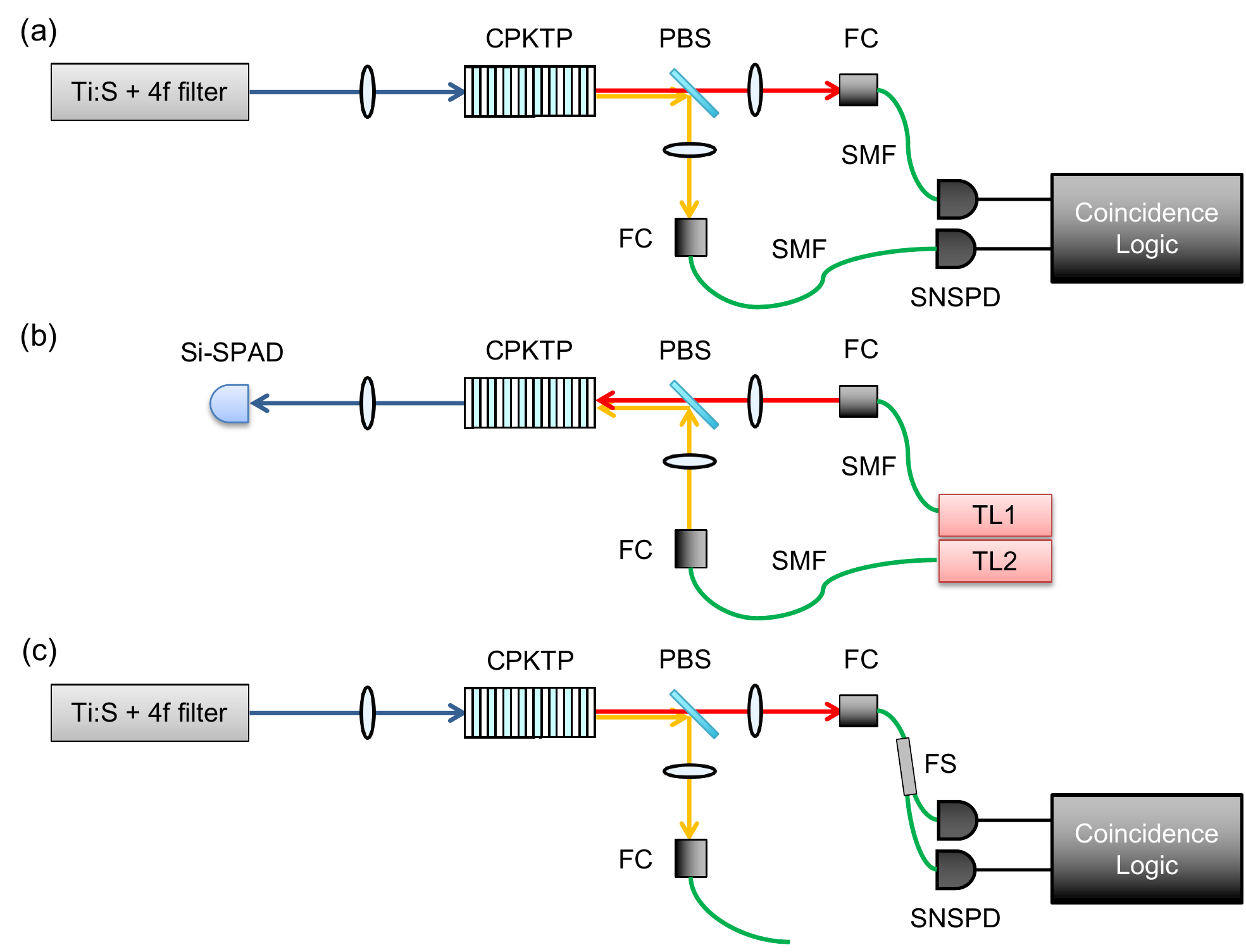}
\caption{Schematic diagram of our experimental setup. (a) Main setup for detection of SPDC photons produced by the CPKTP crystal. (b) Setup for frequency-resolved sum-frequency generation (SFG) measurement to reconstruct a phase-matching function of the CPKTP crystal. (c) Fiber-based Hanbury Brown-Twiss setup for the measurement of the second-order auto-correlation function of the signal mode. PBS, polarizing beamsplitter; FC, fiber coupler; SMF, single-mode optical fiber; SNSPD, superconducting nanowire single-photon detector; TL, tunable laser; FS, fiber beamsplitter. }
\label{setup}
\end{figure}

\begin{figure}[t!]
\centering\includegraphics[width=1\columnwidth , clip]{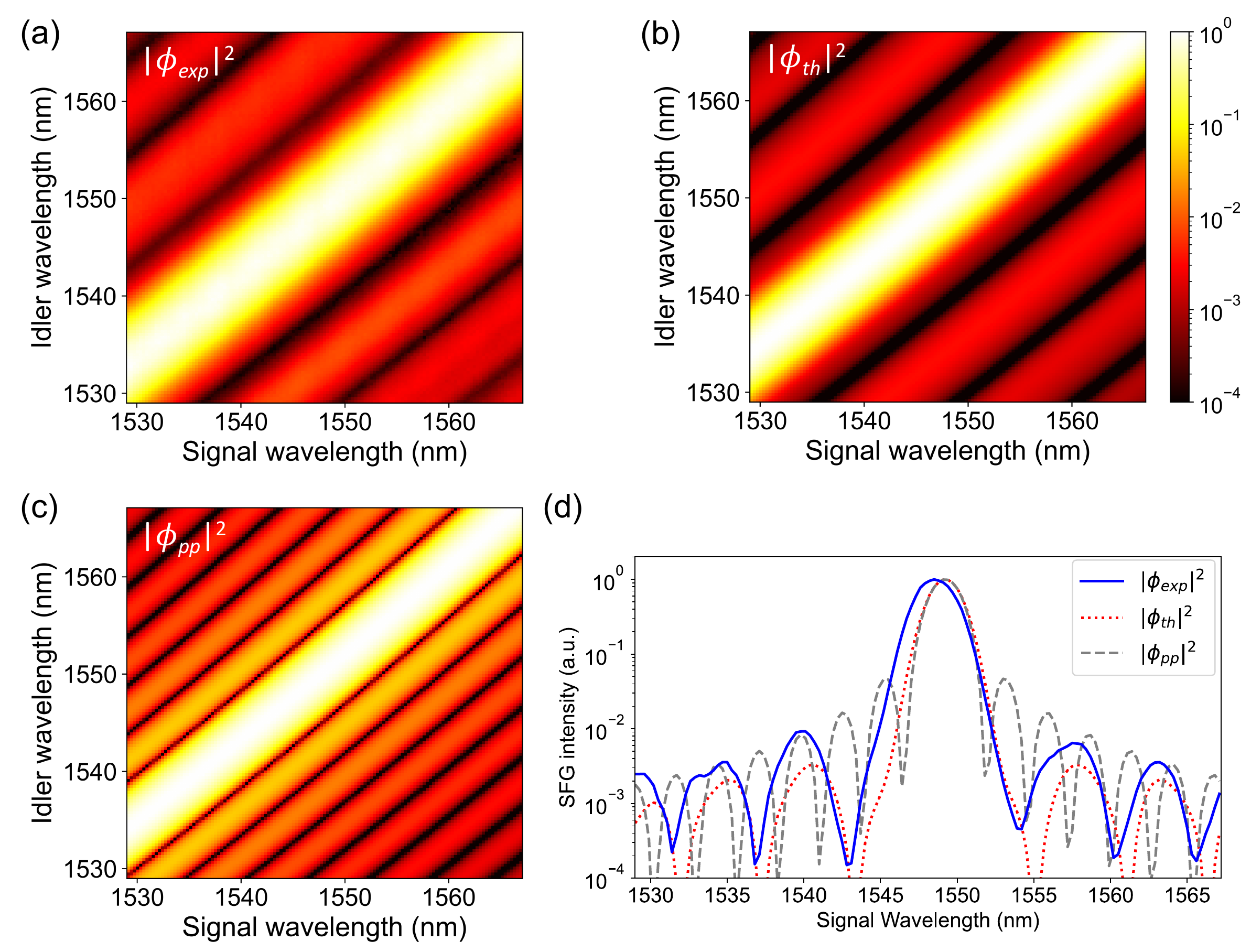}
\caption{Phase-matching functions. (a) CPKTP crystal experimentally observed via frequency-resolved SFG ($|\phi_{\textrm{exp}} (\Delta k (\lambda_s, \lambda_i) )|^2$). (b) Theoretical prediction for the CPKTP crystal ($|\phi_{\textrm{th}} (\Delta k (\lambda_s, \lambda_i) )|^2$). (c) Theoretical prediction for a PPKTP crystal with $L = 10$ mm $(|\phi_{\textrm{pp}} (\Delta k (\lambda_s, \lambda_i) )|^2)$. (d) Cross sectional view of (a)-(c), where the signal and idler wavelengths ($\lambda_s, \lambda_i$) are anti-correlated: $1/\lambda_s = 1/\lambda_p - 1/\lambda_i$, $\lambda_p = 775$ nm. 
%The phase-matching function of our CPKTP crystal has highly suppressed, with large spectral spacing peripheral lobes compared to the ones for the PPKTP crystal. 
}
\label{PMF_result}
\end{figure}

Figure \ref{setup} (a) shows the experimental setup for characterization of our SPDC source. 
Our CPKTP crystal designed as discussed above is fabricated by AdVR inc.    
We used Ti:S laser pulses (central wavelength $\lambda_p = 775$ nm, 4.6 nm FWHM) passed through a 4-f spectral filter as a bandwidth-tunable pump source (0.8-4.6 nm FWHM). 
Signal-idler photon pairs produced at 1550 nm are split by a polarizing beamsplitter (PBS) and then collected by independent single-mode optical fibers (SMF). 
For detecting SPDC photons we used superconducting nanowire single-photon detectors (SNSPDs) with $\sim 50$\% system detection efficiency \cite{Miki.2017}. 
%
%\textcolor{red}{
With the pump power of 50 mW and the pump and photon-collection beam waist radii of 200 $\mu$m and 180 $\mu$m, we observed single and coincidence rates of $1.2 \times 10^5$ cps and $4 \times 10^4$ cps, respectively: 
%}
%
note that the ratio of the single and coincidence rates, i.e., heralding efficiency ($\sim 30$\%) is lower than our theoretical prediction ($> 45$\%) \cite{Kaneda.2016}.   
We believe that this is in part due to rough surfaces of the crystal observed via optical microscope. 
Nonetheless, as will be demonstrated, the imperfect collection is not critical for the proof-of-concept demonstration of the spectral factorability that is essentially determined by the poling structure of our crystal and the time-frequency characteristics of the pump pulses. 
In order to directly measure the phase-matching function of our CPKTP crystal, we performed frequency-resolved sum-frequency generation (SFG) measurement \cite{Kaneda.2020}, using a slightly modified setup as shown in Fig. \ref{setup} (b). 
A pair of tunable continuous-wave lasers (TL1 and TL2, the spectral linewidth of $< 50$ kHz, the center wavelength of 1530-1565 nm) coupled to the collection SMFs of the signal and idler modes are used to produce SFG photons. 
SFG photons are then detected by a Si single-photon avalanche diode (Si-SPAD). 
Since SFG is the reverse process of SPDC, a generation rate of SFG photons is proportional to the same phase-matching function $|\phi (\Delta k (\omega_s, \omega_i))|^2$. 
Thus, the phase-matching function of the CPKTP crystal can be revealed by the spectral distribution of SFG photons that can be measured by scanning the wavelengths of TL1 and TL2. 
For the measurement of the spectral factorability of photon pairs, we performed the second-order autocorrelation function measurement of the signal mode \cite{Christ.2011} with a fiber-based Hanbury Brown-Twiss setup \cite{Brown.1956}, as shown in Fig. \ref{setup} (c).

\section{Result and discussion}

\begin{figure}[t!]
\centering\includegraphics[width=1\columnwidth , clip]{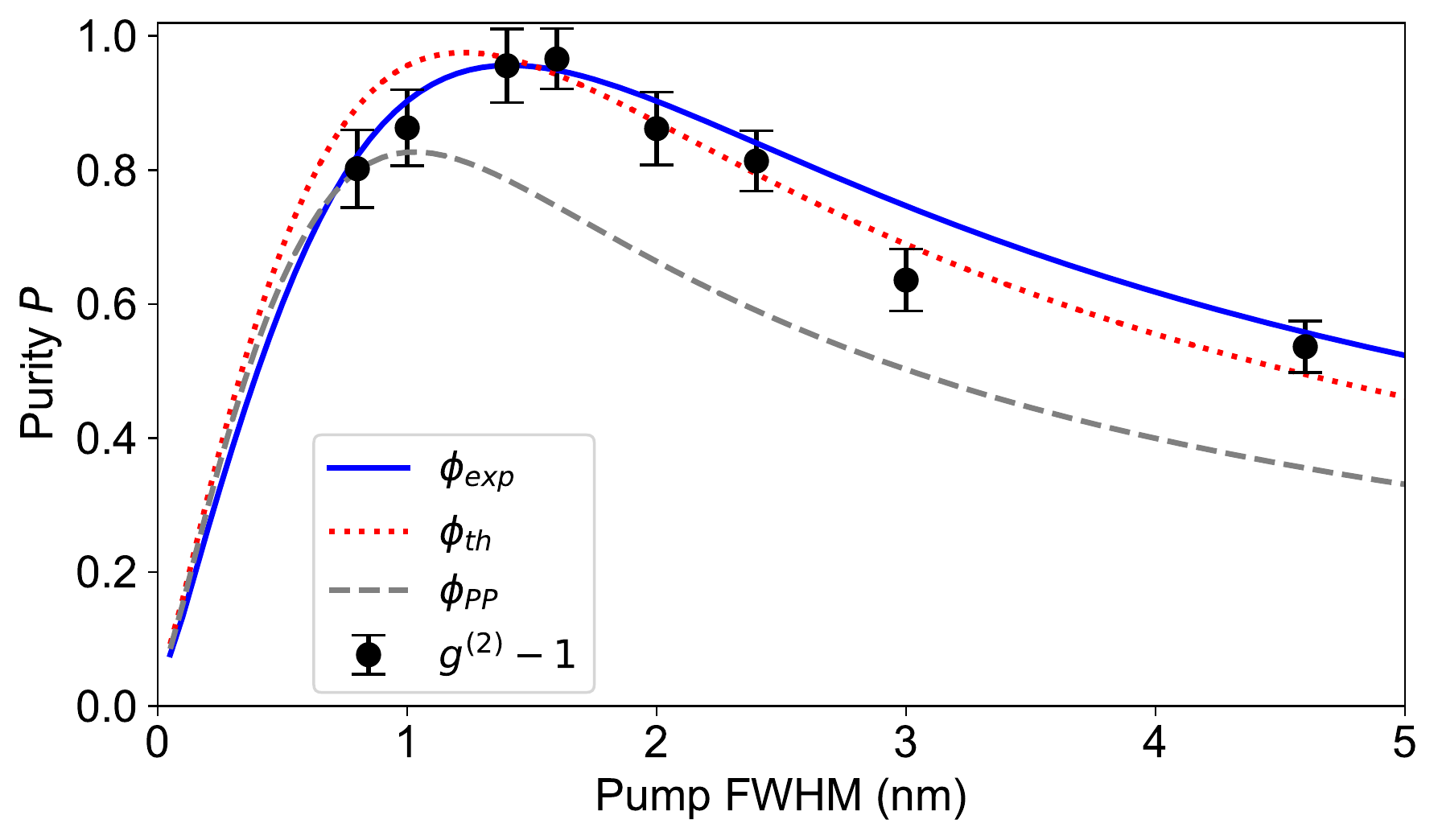}
\caption{Single-photon purity ($P$) versus pump bandwidth (FWHM). $P$ is estimated by a simulated JSA calculated as a product of a numerical Gaussian pump envelope function and the theoretical/experimental phase-matching function. 
The solid, dotted, and dashed curves are respectively for the phase-matching functions $\phi_{\textrm{exp}}(\Delta k (\lambda_s, \lambda_i))$, $\phi_{\textrm{th}}(\Delta k (\lambda_s, \lambda_i))$, and $\phi_{\textrm{pp}}(\Delta k (\lambda_s, \lambda_i))$ shown in Fig. \ref{PMF_result}. 
%We assume that the experimental phase-matching function has no imaginary part and can be written as $\phi_{\textrm{exp}} (\Delta k (\lambda_s, \lambda_i)) = \sqrt{|\phi_{\textrm{exp}} (\Delta k (\lambda_s, \lambda_i))|^2}$. 
Black circles show the measured purity via $g^{(2)}$ measurement ($P = g^{(2)} -1$) for different pump bandwidths. 
The two different methods shown as solid curve and black circles consistently reveal a high purity ($P > 95$\%) for our CPKTP crystal with the optimal pump bandwidth (1.4 nm). 
}
\label{purity_bandwidth}
\end{figure}

Figure \ref{PMF_result} (a) is the observed phase-matching function $|\phi_{\textrm{exp}} (\Delta k (\lambda_s, \lambda_i) )|^2$ for our CPKTP crystal 
(note that all the graphs in Fig. \ref{PMF_result} show the phase-matching \textit{intensity} distribution in \textit{log} scale, while Fig. \ref{concept} and \ref{design} shows the phase-matching \textit{amplitude} distribution in \textit{linear} scale). 
With the high-resolution (0.04 nm), high signal-to-noise ratio ($>40$ dB) SFG measurement, we confirm that $|\phi_{\textrm{exp}} (\Delta k (\lambda_s, \lambda_i) )|^2$ is close to the predicted phase-matching function $|\phi_{\textrm{th}} (\Delta k (\lambda_s, \lambda_i) )|^2$ shown in Fig. \ref{PMF_result} (b), where the Sellmeier equations \cite{Konig.2004} and our poling design (shown in Fig. \ref{design}) are taken into account. 
For comparison,  in Fig. \ref{PMF_result} (c), we also show the theoretical phase-matching function $|\phi_{\textrm{pp}} (\Delta k (\lambda_s, \lambda_i) )|^2$ for a PPKTP crystal ($L = 10$ mm). 
Figure \ref{PMF_result} (d) is cross sectional view of Figs. \ref{PMF} (a)-(c), where the signal and idler wavelengths are anti-correlated: $1/\lambda_s =  1/\lambda_p - 1/ \lambda_i$ ($\lambda_p =  775$ nm).
%7.1 and 8.6 dB suppression each relative to sinc, 7.8 dB in average. 20, 22, and 21 dB suppression relative to the main peak. 
We see that the ratio of the largest peripheral lobe and the main peak of our CPKTP crystal is 20 dB much larger than that for the PPKTP crystal (12 dB). 
%
%Thus, our multi-order QPM technique successfully demonstrated the suppression of the peripheral lobes.  
%
Moreover, the phase-matching function of our CPKTP crystal has a larger spectral spacing between the main peak and lobes. 
As will be discussed, this allows us to use a mild spectral filtering to selectively eliminate the peripheral lobes. 
The slight difference in the center wavelengths of the observed and theoretical phase-matching functions may be caused by the difference of the actual indices of refraction of the CPKTP crystal and the Sellmeier equations. 
We also observed that $|\phi_{\textrm{exp}} (\Delta k (\lambda_s, \lambda_i) )|^2$ has a wider bandwidth of the main peak and larger heights of the lobes compared to $|\phi_{\textrm{th}} (\Delta k (\lambda_s, \lambda_i) )|^2$: this may be due to imperfect poling in some regions of the crystal. 

We then estimated the single-photon spectral purity of our SPDC source. 
The spectral purity $P$ is estimated by Schmidt decomposition \cite{Eberly.2006} of a simulated JSA as a product of the phase-matching function observed by the frequency-resolved SFG measurement and a numerical Gaussian function in Eq. \ref{pump}: $f (\omega_s, \omega_i) = \alpha (\omega_s +\omega_i) \phi_{\textrm{exp}} (\omega_s,\omega_i)$. 
We assume that $\phi_{\textrm{exp}} (\Delta k (\omega_s,\omega_i))$ has no imaginary part and can be written as $\phi_{\textrm{exp}} (\Delta k (\lambda_s, \lambda_i)) = \sqrt{|\phi_{\textrm{exp}} (\Delta k (\lambda_s, \lambda_i))|^2}$. 
Figure \ref{purity_bandwidth} shows the spectral purity versus the pump bandwidth (FWHM). 
Here we also show the purities of JSAs for the theoretical phase-matching functions $\phi_{\textrm{th}} (\Delta k (\lambda_s, \lambda_i))$ and $\phi_{\textrm{pp}} (\Delta k (\lambda_s, \lambda_i))$ of the CPKTP and PPKTP crystals, respectively. 
The maximum purity for $\phi_{\textrm{exp}}(\Delta k (\lambda_s, \lambda_i))$ is $P = 95.6$\% with the pump bandwidth of 1.4 nm, much higher than that for  the PPKTP crystal ($P = 82.7$\%). 
For the theoretical phase-matching function of the CPKTP crystal, the maximum purity is $P = 97.5$\% with 1.2 nm pump bandwidth: 
The slight discrepancy for $\phi_{\textrm{exp}}(\Delta k (\lambda_s, \lambda_i))$ and $\phi_{\textrm{th}}(\Delta k (\lambda_s, \lambda_i))$ is due to the larger peripheral lobes and the wider main peak of our experimental phase-matching function as observed in Fig. \ref{PMF_result}. 
The spectral purity was also characterized by the measurement of the second-order autocorrelation function $g^{(2)}$ for the signal mode \cite{Christ.2011}, shown as black circles in Fig. \ref{purity_bandwidth}. 
We see that the purities obtained by the $g^{(2)}$ measurements ($P = g^{(2)} - 1$) for several different pump bandwidths are in excellent agreement with our estimation based on the observed phase-matching function.
Thus, the two different measurements consistently reveals the high-factorability photon pairs from our CPKTP crystal.  

\begin{figure}[t!]
\centering\includegraphics[width=1\columnwidth , clip]{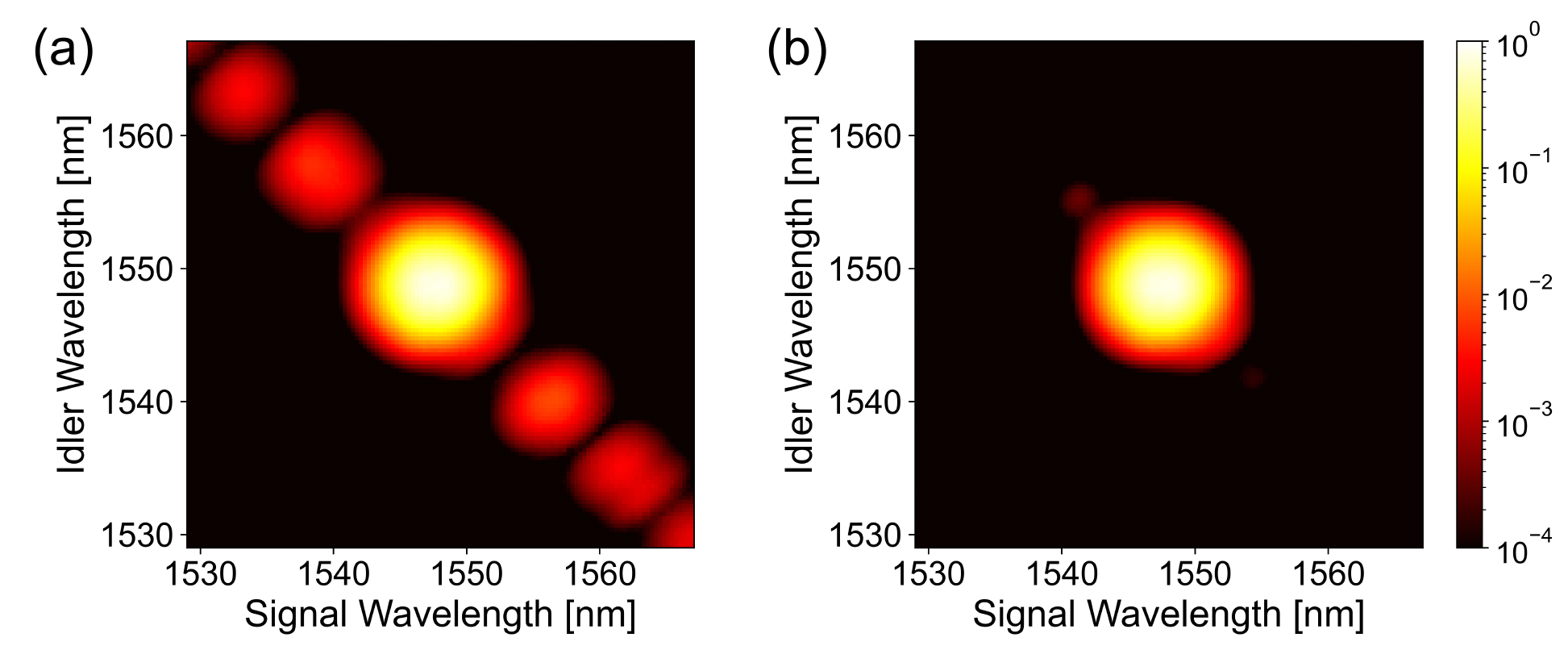}
\caption{Possible enhancement of the spectral factorability via spectral filtering. (a) Predicted JSI with the observed phase-matching function and the optimal Gaussian pump (1.4 nm FWHM). (b) The same JSI as in (a) with 13-nm band flat-top filters applied in the signal and idler modes. The spectral purity is enhanced to $ > 99.9$\% after the spectral filtering while the reduction of the spectral heralding efficiency is only 0.5\%.  }
\label{JSI_filter}
\end{figure}

We note that the spectral factorability of our source can be further improved by mild spectral filtering. 
The predicted JSI with the optimal pump bandwidth (1.4 nm) is shown in Fig. \ref{JSI_filter} (a). 
The peripheral lobes are highly suppressed but still remain, resulting in the limited spectral purity.  
However, thanks to the large spectral spacing of the main peak and the peripheral lobes, one can selectively filter out the peripheral lobes with mild spectral filters. 
Figure \ref{JSI_filter} (b) shows the JSI with applied 13-nm band flat-top filters in the signal and idler modes. 
The spectral purity is enhanced to $P = 99.9$\% with only 0.5\% reduction in the spectral heralding efficiency. 
Thus, our multi-order QPM technique implemented with standard poling technique will be practically useful for producing high-purity, low-loss photon pairs, incorporated with efficient collection techniques of SPDC photons demonstrated in previous works \cite{Christensen.2013,Kaneda.2016,Pickston.2021}. 

\section{Conclusion}
We have demonstrated that the multi-order QPM technique is useful for producing spectrally factorable SPDC photon pairs. 
Our CPKTP crystal satisfying GVM and 1st- to 31st-order QPMs allow to produce photon pairs at the telecom C-band with the positively correlated,  approximate Gaussian phase-matching function. 
Thanks to the highly suppressed peripheral lobes, we observed a high spectral purity ($> 95$\%) of individual SPDC photons via simulation of JSA with the measured phase-matching function and measurement of the second-order autocorrelation function.  
The obtained spectral purity can be further improved to $99.9$\% with a low ($< 0.5$\%) reduction of the spectral-mode heralding efficiency. 
We expect that our source incorporated with optimized photon collection optics and spectral filters will enable to produce low-loss, high-factorability photon pairs, an ideal resource for multi-photon quantum information applications. 

\section*{Funding}
JSPS KAKENHI Grant Number JP18H05949 and JP19H01815, MEXT Quantum Leap Flagship Program (MEXT Q-LEAP) Grant Number JPMXS0118067581, Matsuo Academic Foundation, and Murata Academic Foundation.

\section*{Acknowledgments}
We thank So-Young Baek for helpful discussions. 

\section*{Disclosures}
The authors declare no conflicts of interest.

%\item When datasets are included as integral supplementary material in the paper, they must be declared (e.g., as "Dataset 1" following our current supplementary materials policy) and cited in the DAS, and should appear in the references.

%\bmsection{Data availability} Data underlying the results presented in this paper are available in Dataset 1, Ref. [3].

%\item When datasets are cited but not submitted as integral supplementary material, they must be cited in the DAS and should appear in the references.

%\bmsection{Data availability} Data underlying the results presented in this paper are available in Ref. [3].

%\item If the data generated or analyzed as part of the research are not publicly available, that should be stated. Authors are encouraged to explain why (e.g.~the data may be restricted for privacy reasons), and how the data might be obtained or accessed in the future.

%\section{Data availability} Data underlying the results presented in this paper are not publicly available at this time but may be obtained from the authors upon reasonable request.

%\bmsection{Supplemental document}
%See Supplement 1 for supporting content. 

%%%%%%%%%%%%%%%%%%%%%%% References %%%%%%%%%%%%%%%%%%%%%%%%%

%Add references with BibTeX or manually.
%\cite{Zhang:14,OSA,FORSTER2007,Dean2006,testthesis,Yelin:03,Masajada:13,codeexample}

%%%%%%%%%% If using BibTeX:
\bibliography{CPKTP_reference02.bib}
%\input{CPKTP02_mod.bbl}
%\end{backmatter}

\end{document}